# When Finance Meets Physics:
# The Impact of the Speed of Light on Financial Markets and their Regulation

by

James J. Angel, Ph.D., CFA
Associate Professor of Finance
McDonough School of Business
Georgetown University
509 Hariri Building
Washington DC 20057 USA

1.202.687.3765
angelj@georgetown.edu

*The Financial Review*, Forthcoming, May 2014 · Volume 49 · No. 2

Abstract:

Modern physics has demonstrated that matter behaves very differently as it approaches the speed of light. This paper explores the implications of modern physics to the operation and regulation of financial markets. Information cannot move faster than the speed of light. The geographic separation of market centers means that relativistic considerations need to be taken into account in the regulation of markets. Observers in different locations may simultaneously observe different "best" prices. Regulators may not be able to determine which transactions occurred first, leading to problems with best execution and trade-through rules. Catastrophic software glitches can quantum tunnel through seemingly impregnable quality control procedures.

The author is also on the board of directors of the Direct Edge stock exchanges (EDGX and EDGA). I wish to thank the editor and referee for extremely helpful comments and suggestions. I also wish to thank the U.K. Foresight Project, for providing financial support for an earlier version of this paper. All opinions are strictly my own and do not necessarily represent those of Georgetown University, Direct Edge, the U.K. Foresight Project, *The Financial Review*, or anyone else for that matter.

**Introduction**

Our markets move much faster than ever before. In a few short years, we have gone from a world in which humans traded face to face with humans to one in which computers trade with computers. Rather than responding in human-scale time, our markets now respond in computer-scale time. Traders fret about whether other traders have a millisecond advantage over them. Firms pay extra in order to have their computers co-located in the same data centers as stock exchanges, so that their trades are not delayed by the length of time it takes for an electronic signal to get from their office to the exchange.

What does this speed mean for our financial markets? In the early 20th century, physicists such as Einstein (1905) discovered that matter behaves differently at speeds approaching the speed of light than it does at lower speeds. Do our financial markets also behave differently as they approach trading at the speed of light? Are markets at the speed of light different? Just as intuitions gained from low-speed Newtonian mechanics need to be modified near the speed of light, intuitions – and regulations -- gained from low-speed markets also need to be modified as trading approaches the speed of light.

**I  Special relativity and modern physics**

Market participants engage in a wide variety of trading strategies, some of which involve reacting quickly to market conditions and trading rapidly. These so-called "high frequency" traders are often implementing some very old and fairly low-tech strategies. Examples of high-frequency strategies include arbitrage, market making, and reacting to news. Because these traditional trading strategies are rather simple and easy to duplicate, traders face intense competition in their implementation. They race to snap up profitable trading opportunities before they disappear. An arbitrageur who comes in second in the race for a profitable trade still loses, whether by one minute, one second, or one nanosecond. For this reason, such traders invest heavily to make sure that their trading systems respond as fast as possible to market conditions. Exchanges have sped up their response times and major exchanges now have "latencies" of less than one millisecond (one thousandth of a second).[1] Exchanges and practitioners now routinely time stamp their messages to the millisecond or nanosecond.[2]

However, speed issues are now bumping up against physical issues. These physical issues are not just computing power, but actual limitations from physics itself. For example, the famous Michelson and Morley (1887) experiments demonstrated that the speed of light is a constant, regardless of whether the light source is moving towards the observer or away from the observer. This invariance of the speed of light has some important implications, as shown by Einstein (1905) in his famous paper showing the equivalence of mass and energy in the equation $e = mc^2$. Because nothing can move significantly faster than light in a vacuum, the speed of light thus becomes a speed limit on the transmission of information as well.[3]

The limitations based on the speed of light are affecting financial markets in a variety of ways. For example, some firms pay extra to co-locate their servers in stock exchange data centers so that their orders can get to the exchange's matching engine sooner. In one millisecond, light travels about 300 km, roughly the distance between Boston and New York. (In real computer networks, the signal travels a little slower because of the delay in going through a solid and because of switching delays in the network.) In one microsecond (one millionth of a second), light travels about 300 meters, or just over 3 (American) football fields. In one nanosecond (one billionth of a second), light travels about 30 cm, or about a foot. In a very close race between two computerized traders, that nanosecond just might make the difference between catching a profitable trade and missing it.

---

[1] For example, NasdaqOMX reports on its web page average latencies of less than 100 microseconds, or 0.1 millisecond. http://www.nasdaqtrader.com/Trader.aspx?id=Latencystats, Accessed February 15, 2013.

[2] Not all high-speed strategies are traditional old strategies, however. Clever traders can make use of the differences of the modern environment to find new trading opportunities. Ding, Hanna, and Hendershott (2014) documents that market participants can use the proprietary high speed data feeds from the exchanges to construct their own version of the National Best Bid and Offer (NBBO) that is slightly faster than the official NBBO. This can give rise to trading strategies designed to exploit these discrepancies, such as by picking off orders in dark pools that use the slower official NBBO as a reference price (Ding, Hanna, and Hendershott, 2014).

[3] This invariance of the speed of light also has implications for the concept of time. One of the interesting findings of special relativity is that as an object moves faster time slows down. This leads to the famous "twins paradox" in which twins born on the same day age at different rates when one travels close to the speed of light. This has been demonstrated experimentally by taking very accurate atomic clocks on high-speed jet rides in different directions. See J.C. Hafele and R. E. Keating (1972a,b). Furthermore, as Pogge (2009) shows, GPS units on earth need to adjust for fact that time appears to flow at different rates at different altitudes due to the effects of gravity on light, as predicted by Einstein (1908). See Laughlin, Aguirre, and Grundfest (2014) for a discussion as it relates to GPS, tower heights, and the speed of trading.

**II Regulatory concerns about high-speed markets**

Modern physics has made it clear that matter and time behave differently as speed approaches the velocity of light. We can use the insights of physics to see what pitfalls may occur in the understanding and regulation of high-speed markets. As noted by Aspinwall (1993) and others, financial market regulation has multiple – and often conflicting – objectives, such as consumer protection, fairness, resource allocation, economic efficiency, capital formation, soundness of financial institutions, and economic stability.

*Consumer protection*

Public policy decisions in the United States and Europe have resulted in competitive equity markets. As the old monopoly markets privatized, public policies such as Regulation NMS in the United States and MiFID in Europe adopted pro-competition policies to prevent the economic problems resulting from for-profit monopolies. New exchanges and multilateral trading platforms have arisen, resulting in a much more competitive trading environment. Brokerage firms also provide competition by trading directly with their customers rather than sending orders to other exchanges. These changes have resulted in equity market networks that consist of geographically diverse trading nodes. While Angel, Harris, and Spatt (2011) note that this combination of competition and technology has generally resulted in improvements in measures of market quality, they also express concerns about the technological stability of current U.S. market structure and the distortions caused by so-called maker-taker pricing. Other critics such as Arnuk and Saluzzi (2012) and Bodek (2013) contend that the current market structure provides unfair advantages to high speed traders.

*Best execution and trade through rules*

In a world of competing trading venues, some clients have difficulty determining whether their brokers have done a good job executing their trades. Regulators seek to protect investors by requiring that brokers seek "best execution."[4] For example, the United States' "trade through" rule of Regulation NMS (SEC Rule 611, 17CFR242.611) requires exchanges to not "trade through" the quotes of another exchange. An exchange is not permitted to execute a trade at one price when a better price is available on another exchange.

Enforcing such rules in a high-speed world runs into two problems. The first problem stems from the fact that information does not travel faster than light. Because information can only travel at the speed of light, two events that appear to happen at the same time to one observer may appear to happen at different times to a second observer in another reference frame. If two exchanges are geographically separated, the information that one exchange has a better price may not have reached the other exchange before it executes a trade.[5] For example, suppose two exchanges are both offering to sell stock at $10, so both display an ask quote of $10. Then the first exchange receives an order to sell stock at $9 and updates its ask quote, which is broadcast to the market. At the instant the new ask quote is broadcast, but before the broadcast signal reaches the second exchange, observers at the two separate exchanges will observe

---

[4] The Securities and Exchange Commission (2011) states: "Brokers are legally required to seek the best execution reasonably available for their customers' orders. … Some of the factors a broker must consider when seeking best execution of customers' orders include: the opportunity to get a better price than what is currently quoted, the speed of execution, and the likelihood that the trade will be executed." As Macy and O'Hara (1997) point out, this duty stems from common law fiduciary obligations, not specific laws or regulations. Because this duty involves an entire vector of attributes of trade quality, not just price, it is "unwieldy and unenforceable." The European Commission (2004) is quite explicit in Article 21 of its Markets in Financial Instruments Directive (MiFID): "Member States shall require that investment firms take all reasonable steps to obtain, when executing orders, the best possible result for their clients taking into account price, costs, speed, likelihood of execution and settlement, size, nature or any other consideration relevant to the execution of the order."

[5] Wissner-Gross, A.D, and C.E. Freer (2010) examine the implications of information propagation delays to calculate the optimal intermediate locations for conducting arbitrage activity between geographically separate trading locations.

two different "best" ask quote prices. An observer at the first exchange sees the best quoted ask price as $9 and an observer at the second exchange sees the best quoted ask price as $10. Alas, the second exchange receives an order to buy (which will execute against the ask quote) before the updated information that there is a better offer at the other exchange arrives, perhaps due to speed of light issues. The second exchange then fills the order at its ask price of $10. From the second exchange's perspective, it has given the customer best execution. However, an observer at the first exchange would say that the second exchange has traded through the first exchange's ask quote and has violated the trade-through rule to the detriment of the investing public.

How should a regulator determine whether an exchange has broken a "trade-through" rule or not in a high speed environment? One regulatory approach would be to examine the time stamps of the market events, and correct for the cone of information by reconstructing the information environment at each exchange at the time of each trade. This approach raises the second problem in enforcement, in part due to physics. There are limits to the practical precision which can be obtained at reasonable cost – or even any cost - by market participants and regulators. Syncing to a standard time source such as an atomic clock sounds simple in theory, but there are limits to the practical accuracy of what can be implemented, as noted by Kopetz and Ochsenreiter (1987). Even with access to a signal from a standard time source, there are still random propagation delays within the processing hardware that can limit the precision of the resulting time stamps. While Laughlin, Aguirre and Grundfest (2014) claim that at least some market participants are now synchronized at the millisecond level to a standard time source, even clock synchronization at the millisecond level leads to a fundamental uncertainty when latencies are measured in microseconds.

Thus, a regulator attempting to reconstruct the sequence of events and determine whether one exchange filled the order before or after the other exchange displayed a better price may never be able to accurately determine what happened. Such controversies are already occurring, as seen in the debate between Nanex (2013) and Virtu Financial (2013) over whether Fed data were leaked to high-speed traders prior to a Fed announcement. This debate centers on the accuracy of the relevant time stamps and the time needed for signals to travel between Washington DC, New York, and Chicago.

In a similar vein, some have called for time priority to be preserved across different trading venues: they believe that the first to post a quote at a particular price should get the first opportunity to be filled. This idea is simple to enforce within a single exchange's computer. However, enforcing such a rule across geographically separate platforms runs into the same problem as a trade-through rule. Because information can only travel at the speed of light, the current state of the market will appear differently in different geographic locations. A regulator attempting to determine whether a trading platform is obeying the rule has an irreducible uncertainty about the where the market is at a particular point in time.

These issues are similar in some ways to the Heisenberg uncertainty principle, which shows that there is a limit to the precision with which the location and momentum of a particle can be determined. As the best bid is the maximum of the set of geographically separated bids from separate exchanges, an observer at one location may observe a different best bid than an observer at a different location due to the inescapable propagation delays. There is thus an inescapable lower bound on the time precision with which the NBBO can be determined. This physical limitation makes monitoring and enforcement of such a rule problematic at the speeds with which modern financial markets operate.

There are similar, but not identical, questions of uncertainty in quantum mechanics, another realm of modern physics where interesting things happen that are very different from what we commonly observe in the macro-world. A particle could be in different locations at the same time, just with different probabilities. Not only is the location of a particle a probability function, so are other states of a quantum system. A quantum system may exist as a "superposition" of various states, and only when the system is observed does the system become one state or another. Schrődinger (1935) described the famous "Schrődinger's cat" thought example in which a cat is sealed in a box and its life is dependent on the state of a quantum mechanical system which may exist as a superposition of different quantum mechanical states. The cat is both alive AND dead until someone opens the box to observe the state of the system. It is only then that the state of the system – and the cat's life -- is determined.

In some sense, the issue underlying traders' complaints about "phantom liquidity" -- in which liquidity disappears when they attempt to trade against it (Themis Trading, 2012) -- is similar to the issue in "Schrődinger's cat": a market participant does not really know if a published "firm" quote is actually alive (can be traded upon) or dead without actually submitting an order to trade against it. Only once they submit an order will they find out whether their order is filled or whether someone else beat them to it or the quote was cancelled before their order arrived. Traders may therefore feel that they were "traded through" or that they did not get the best execution possible.

*Fairness*

Some critics claim that it is unfair that some firms are allowed to "co-locate" their computers in stock exchange data centers. This proximity to the exchange matching engines allows them to receive market data faster and to submit orders faster than other participants located farther away. However, this power is available to anyone who wishes to pay for it. Angel and McCabe (2012) argue that as long as such services are available on fair and non-discriminatory terms to all market participants, there is nothing especially unfair about the practice of co-location. For centuries brokerage firms have located offices as close as possible to the exchange so that they could get their orders into the exchange faster. The only difference is that the orders are now submitted via electrons and not runners.

Is it fair that some participants have the resources to spend on co-location that others don't? Angel and McCabe (2012) argue that it is no more unfair than the fact the some investors are endowed with more resources to spend on fundamental research, or better brains for finding good investments. If the practice of co-location were banned, the co-locators would merely move to another location as close as possible to the exchange data center. Rather than the exchange getting the revenue, the landlord of the other location would collect more rent. Indeed, some exchanges such as Direct Edge do not offer direct co-location services, but their location in commercial data centers makes it easy for market participants to effectively co-locate by locating very closely in the same commercial data center.[6]

*Economic efficiency*

Some high-frequency trading strategies such as arbitrage and market making clearly help the market to operate more efficiently. News reaction strategies help the market to incorporate information faster. On the other hand, there is an arms race among communication vendors to provide ever faster and more expensive links between data centers.

One of the primary "places" where the limitation due to the speed of light can be seen is in the information transmission between the futures market in Chicago and the equity markets in the New York/New Jersey area. These locations are about 730 miles apart and even in a vacuum it would take light (and therefore information) just under 4 milliseconds to go from one market to the other. In an examination of the effects of the speed of light on transmissions between the futures markets in Chicago and the equity markets in New Jersey, Laughlin, Aguirre, and Grundfest (2014) show the proliferation of microwave networks and their effects on reducing the transmission time between these markets to close to the theoretical speed of light limit of 3.93 milliseconds. They also show that changes in the limit order book quotes at the CME in Chicago result in trades on the equity markets in New Jersey near the theoretical limits of the speed of light and these microwave arrays. Laughlin, Aguirre, and Grundfest (2014) also estimate that these microwave arrays have considerable construction and operating costs and the speed benefit is nearing its theoretical limit.

It is questionable as to the benefits to society of expending resources to squeeze another microsecond out of the time to transmit a signal from the CME's data center in Aurora, IL to NasdaqOMX's data center in Carteret, NJ. Harris (2012) proposes that exchanges delay the processing of orders by a small but random time, which would reduce the incentive to spend resources to be first in line.

---

[6] The author serves on the boards of directors of the Direct Edge stock exchanges (EDGX and EDGA).

Budish et al. (2013) suggest periodic batch auctions which would similarly reduce incentives to spend more money to get to the front of the line. However, it is a matter for future analysis as to whether any actions to slow this arms race would have net benefits to society, or worse, serious negative unintended consequences.

*Soundness of financial institutions*

One of the lessons of quantum physics is that occasionally extreme events can and do occur. Particles can "quantum tunnel" through a solid barrier even when they do not have enough energy to get through it by classical calculations. There is a nonzero probability that a particle such as an electron or even a basketball can get through a solid barrier no matter how high or solid the barrier. Quantum physics therefore suggests that "freak events" will occur.

In our highly complex and nonlinear market network, "freak events" will also occur. For example, freak events can be triggered from programming glitches in which seemingly innocuous minor changes lead to a major catastrophe, which has already happened: on August 1, 2012, Knight Capital Group (2012) nearly failed following a $440 million software glitch. Such catastrophic freak failures would have been caught earlier when every order passed through human hands. Regulators need to be vigilant about such risks. No matter how good the quality control procedures in the financial industry are, sooner or later another catastrophic glitch will seemingly "quantum tunnel" its way into existence. Practitioners and industry need to have contingency plans in place for containing the damage when such events happen.

*Economic stability*

Some high frequency trading strategies, particularly market making strategies, are usually stabilizing to the financial markets. For example, market making provides buying power when others want to sell and selling power when others want to buy. This negative feedback tends to stabilize prices. Arbitrage keeps the prices of related instruments in the proper alignment. However, other activities can destabilize the market. Momentum traders tend to buy when stocks are going up and sell when they are going down, a positive feedback which can accelerate volatile price moves.

Under the right combination of events, the complex market network of interconnected exchanges and off-exchange participants can become unstable, as demonstrated by the US experience of the "Flash Crash" on May 6, 2010.[7] On a day of heavy trading volume and generally falling prices, a large investment firm placed a large sell order on futures contracts tied to the S&P 500 and the price of the contract plummeted. There were reports of data problems at several firms. Several high frequency trading firms withdrew from trading, citing "data integrity" problems. Given their need to respond quickly to market conditions, their reliance on accurate and timely market data, and their generally thin profit margins, these firms felt it was only prudent for them to withdraw from the market when they suspected problems with their data feeds. The removal of a significant amount of market making and arbitrage activity, while other investors continued to trade, caused highly erratic cash equity prices. The market quickly recovered, but over 20,000 "clearly erroneous" trades were cancelled. Even under normal circumstances, there are concerns that high-speed trading may exacerbate market volatility. At various times the various computer programs may interact in such a way as to cause a temporary shortage of liquidity which would increase volatility.

**Conclusions and Summary**

The speed of our financial markets has accelerated in recent years as we have moved from human-intermediated markets to machine-intermediated markets. Furthermore, our markets are now geographically diverse complex networks of competing trading platforms. Modern physics has taught us that objects moving at high speed behave differently from objects moving at low speed. The speed of light provides an upper limit on the speed with which information can flow. This has large implications for any rules that are based on prices at a given time. Two observers in two different locations can simultaneously observe two different "best" prices.  Similarly, the inherent limits to the precision of time measurement mean that there will be an irreducible uncertainty as to when particular events occurred.  Again, this will have serious implications for implementation and enforcement of rules seeking to ensure that a customer gets the best price available at a given time.

---

[7] See the SEC/CFTC (2010a,b) joint reports and Kumar et al. (2012) for more information on the Flash Crash.

The highly nonlinear nature of the financial market network is also cause for concern.  Under normal conditions many financial markets participants provide negative feedback that tends to stabilize the markets as they buy low and sell high.  Others, such as momentum investors add positive feedback to the market.  Conditions can occur that lead to temporarily destabilizing positive feedback.  As trading speed increases, physical issues and lessons from physics become more important for regulators and practitioners alike.